\documentclass[rnote]{aa}
\usepackage{graphicx}
\usepackage{txfonts}
\usepackage{natbib}
\bibpunct{(}{)}{;}{a}{}{,} 
\begin{document}
\title{Near-infrared spectro-interferometry of three OH/IR Stars
with the VLTI/AMBER instrument
\thanks{Based on observations made with the ESO VLT Interferometer
at the La Silla Paranal Observatory under programme ID 081.D-0325.}} 
\author
{A.~E. Ruiz-Velasco\inst{1,2}\and
M. Wittkowski\inst{1}\and
A. Wachter\inst{2}\and
K.-P. Schr\"oder\inst{2}\and
T. Driebe\inst{3,4}
}
\institute{
ESO, Karl-Schwarzschild-Str. 2,
85748 Garching bei M\"unchen, Germany,
\email{alma@astro.ugto.mx}
\and
Departamento de Astronom\'ia, Universidad de Guanajuato.
Apartado Postal 144, 36000 Guanajuato, Mexico.
\and
Max-Planck-Institut f\"ur Radioastronomie, Auf dem H\"ugel 69,
53121 Bonn, Germany
\and
German Aerospace Center, Space Agency, 
K\"onigswinterer-Stra\ss{}e 522-524, D-53227 Bonn, Germany
}
\date{Received \dots; accepted \dots}
\abstract{}
{We investigate the molecular and dusty environment of OH/IR stars
in order to characterize the mass-loss process during the 
tip-AGB superwind phase.}
{Employing the AMBER instrument at the VLT Interferometer we obtained 
near-infrared $H$- and $K$-band spectro-interferometric 
observations of the three OH/IR stars \object{IRAS~13479-5436},
\object{IRAS~14086-6907} and \object{IRAS~17020-5254} with a 
spectral resolution of about 35. We use a two-component geometrical model,
consisting of a uniform disk and a Gaussian disk, to obtain
characteristic angular sizes of the central stellar sources and their 
dust envelopes, as well as the flux ratios between these components.}
{Angular uniform disk diameters of the three central components of the 
objects above have values between 3.2\,mas and 5.4\,mas. 
For their dust envelopes, we find FWHM values between 17.1\,mas and
25.2\,mas. The three objects show significantly different flux contributions 
of the shells to the total near-IR flux of  61\%, 38\%, and 16\% for
IRAS~13479-5436, IRAS~14086-6907, and IRAS~17020-5254, respectively.
According to distance estimates from the literature, the central stellar 
components have radii between 900\,R$_\odot$ and 1400\,R$_\odot$, 
while their dust envelopes reach FWHM values between 9000\,R$_\odot$ and 13000\,R$_\odot$.
The visibility functions of all three sources exhibit
wavelength variations that resemble those of earlier VLTI/AMBER observations of semi-regular
and Mira variable AGB stars. These are interpreted as characteristic
of atmospheric molecular layers lying above the photosphere.}
{The derived characteristic sizes of both, the central stellar atmospheres and 
dust envelopes are consistent with the canonical properties of OH/IR stars.
The spectral visibility variations resemble those of other AGB stars and 
indicate the presence of molecular layers, confirming that these are a 
common phenomenon among AGB stars of very different 
luminosities and mass-loss rates, alike. We also find that the dust envelopes
have a clearly larger optical depth than those known for Mira stars.
We interpret this as an expected result of the ``superwind'' phase, the 
final 10\,000 to 30\,000 years of AGB-evolution, when the mass-loss rate 
increases by a factor of 10-100. By their different optical depths, the 
three dust shells studied here may represent different stages of the
``superwind'' and different initial masses.}
\titlerunning{Near-IR spectro-interferometry of OH/IR stars}
\maketitle
\section{Introduction}
OH/IR stars are medium-mass, highly evolved objects, which either have nearly 
reached the tip of the asymptotic giant branch (AGB) or may just have
left the AGB toward the post-AGB stage. Observational 
characteristics are an infrared excess and the presence of hydroxyl 
(OH) masers, as well as variability due to the release of cool, 
dust-driven winds with a high mass-loss rate (``superwind''), prior
to forming a planetary nebula (PN). These critical, late phases are 
of fundamental interest to a complete understanding of stellar evolution.

Mass-loss rates of stars at the tip of the AGB can reach values of up to
$\sim10^{-4}M_{\sun}$\,yr$^{-1}$ \citep{beck10}. Such a ``superwind'' 
can expel a considerable fraction of the stellar mass in a relatively short 
time ($<$100\,000 years). Time-dependent, dust-driven wind models 
\citep{wachter02,mattsson10} reproduce such mass-loss rates reasonably 
well for carbon-rich AGB stars. A pulsating photosphere for the initial 
mechanical energy input 
and low effective temperatures lead to conditions favorable for dust 
formation in the denser regions of such a circumstellar envelope (CSE). Due
to their high absorption coefficient, carbon dust grains are accelerated by the
stellar radiation such that the star is eventually surrounded by an 
expanding envelope. It significantly changes the spectral appearance of 
the central star, due to its large optical depth at visual and near-IR 
wavelengths and thermal re-emission in the farther IR.
Except for the case of the dust-enshrouded Mira variable IRC -20197
\citep{jeong03}, such self-consistent wind models do not yet exist for oxygen-rich
AGB stars, and, moreover, recent studies by \citet{woitke06} and
\citet{hoefner07} indicated fundamental problems
in our understanding of the wind-driving mechanism for 
oxygen-rich AGB stars due to an insufficient radiation pressure 
on silicate grains.

However, previous high spatial resolution work on OH/IR stars
confirmed the presence of circumstellar dust shells of large optical depth
for oxygen-rich stars near the tip of the AGB.
$K$-band speckle interferometry of the OH/IR star OH 104.9+2.4 \citep{riechers04}
measured a circumstellar dust shell of large optical depth
of 6.5 at 2.2\,$\mu$m with an apparent angular diameter of $47\pm3$~mas (FWHM),
corresponding to $112\pm13$~AU. 
\citet{chesneau05} resolved the mid-IR dusty envelope of the OH/IR star
OH 26.5+0.6 to a FWHM size of 286\,mas $\times$ 214\,mas, or 
about 390 AU $\times$ 290 AU.
Size and complexity of the circumstellar structures 
of OH/IR stars narrate the mass-loss history of the final AGB stages,
which in turn shapes the future planetary nebula.
The recent development of very high angular resolution instrumentation 
has finally enabled us to decipher this crucial evidence. 

In this research note, we study the three OH/IR stars
\object{IRAS 13479-5436}, \object{IRAS 14086-6907} and 
\object{IRAS 17020-5254}. 
All three objects exhibit a typical two-peak maser profile at 
1612 MHz \citep{lintel91}, which is indicative of an expanding CSE.
Furthermore, a large infrared excess and a 
strong $9.7\mu$m silicate emission feature in the mid-IR spectra
indicate the presence of an optically thick, dust-rich CSE.
Table~\ref{table:colors} shows the near-infrared photometry and colors 
of \citet{fouque92}, the bolometric distances given by \citet{lepine95}, the average
flux densities $S_{\rm OH}$ at 1612 MHz between the two OH maser peaks,
and the respective expansion velocities $v_{\rm exp}$ of the OH shells 
according to \citet{lintel91}. 

\begin{table}
\caption{Characteristics of the sources}
\label{table:colors}
\centering
\begin{tabular}{l c c c c c c c c} 
\hline\hline
IRAS Name         & $K$ & $J-K$ & $K-L$ & D  & $S_{\rm OH}$ & $v_{\rm exp}$  \\ 
              &         &       &       &  (kpc) &  (Jy)  &  (km s$^{-1}$) \\
\hline
13479-5436 & 4.89 & 4.39 & 2.15 & 2.9 & 0.68 &  17.7 \\
14086-6907 & 4.27 & 3.28 & 1.98 & 2.4 & 6.15 &  13.4 \\
17020-5254 & 4.15 & 2.59 & 1.38 & 2.5 & 1.08 &  13.4 \\
\hline
\end{tabular}
\end{table}
%
\begin{table*}
\caption{Observation log of 4 April 2008}
\label{table:log}
\centering
\begin{tabular}{l c c c l c c c}
\hline\hline
Target & R.A & Dec & Starting time & Category & Projected baseline & P.A.  & Seeing  \\
       &     &     &               &          & (m)         & (deg) & (arcsec) \\\hline
IRAS 13479-5436 & 13 51 12.6 & -54 51 13 & 05:15:18.650 & Science    & 15.89/31.76/47.65 & -110 & 0.83 \\
HD 126284        & 14 26 22.2 & -55 15 33 & 05:39:33.025 & Calibrator &  &  & 1.01  \\
IRAS 14086-6907 & 14 12 50.5 & -69 21 10 & 06:05:21.233 & Science    & 15.49/30.97/46.47 & -100 & 0.86 \\
HD 114912        & 13 15 25.5 & -69 40 45 & 06:33:44.509 & Calibrator &  &   & 1.47 \\
IRAS 17020-5254 & 17 06 00.7 & -52 58 47 & 06:58:34.476 & Science    & 15.92/31.83/47.76 & -128 & 0.96 \\
HD 154486        & 17 08 08.2 & -48 53 01 & 07:24:43.020 & Calibrator    &   &   & 0.95 \\
\hline
\end{tabular}
\end{table*}
\section{Observations and Data Reduction}
We obtained near-infrared $H$- and $K$-band spectro-interferometric
observations of the three OH/IR stars \object{IRAS 13479-5436},
\object{IRAS 14086-6907}, and \object{IRAS 17020-5254}, using the AMBER
instrument \citep{petrov07} of the VLT Interferometer (VLTI)
in low resolution mode ($R\sim35$) and three of the 1.8~m Auxiliary
Telescopes (AT). These were positioned on stations E0, G0 and H0, 
providing ground baselines of 16\,m, 32\,m and 48\,m, along the same 
ground position angle of -109\,\degr\  East of North.
The AMBER low resolution data also include coverage of the $J$ band, 
but because of low J-band fluxes and, hence the poor quality of the 
respective visibility data, we were unable to include the J-band 
in our study.

Because our target stars are dust-enshrouded objects, they have
$V$ magnitudes too faint to use them as Coud{\'e} guide stars. Instead, 
we resorted to the off-axis guide-star option of the VLTI and chose the 
stars GSC\,0866800251, GSC\,0924401731, and GSC\,0872601051 from the 
HST Guide Star Catalog II \citep{lasker08}. These guide stars are located at
distances of 42\,arcsec, 39\,arcsec, and 55\,arcsec, respectively, from our
target stars listed above. The external VLTI fringe tracker FINITO was not 
used. The AMBER detector integration time (DIT) was set at 0.1\,sec.
After each target a calibration star of well known diameter 
from the catalog of \citet{merand05} was observed (HD 126284 with
$\Theta_\mathrm{UD}=1.14 \pm 0.02$\,mas, HD 114912 with 
$\Theta_\mathrm{UD}=0.95 \pm 0.01$\,mas, and HD 154486 with
$\Theta_\mathrm{UD}=1.06 \pm 0.02$\,mas).
The observations presented here were all obtained on 4 April 2008. 
Table~\ref{table:log} summarizes the observational details.

\begin{figure}
\centering
\resizebox{1\hsize}{!}{\includegraphics{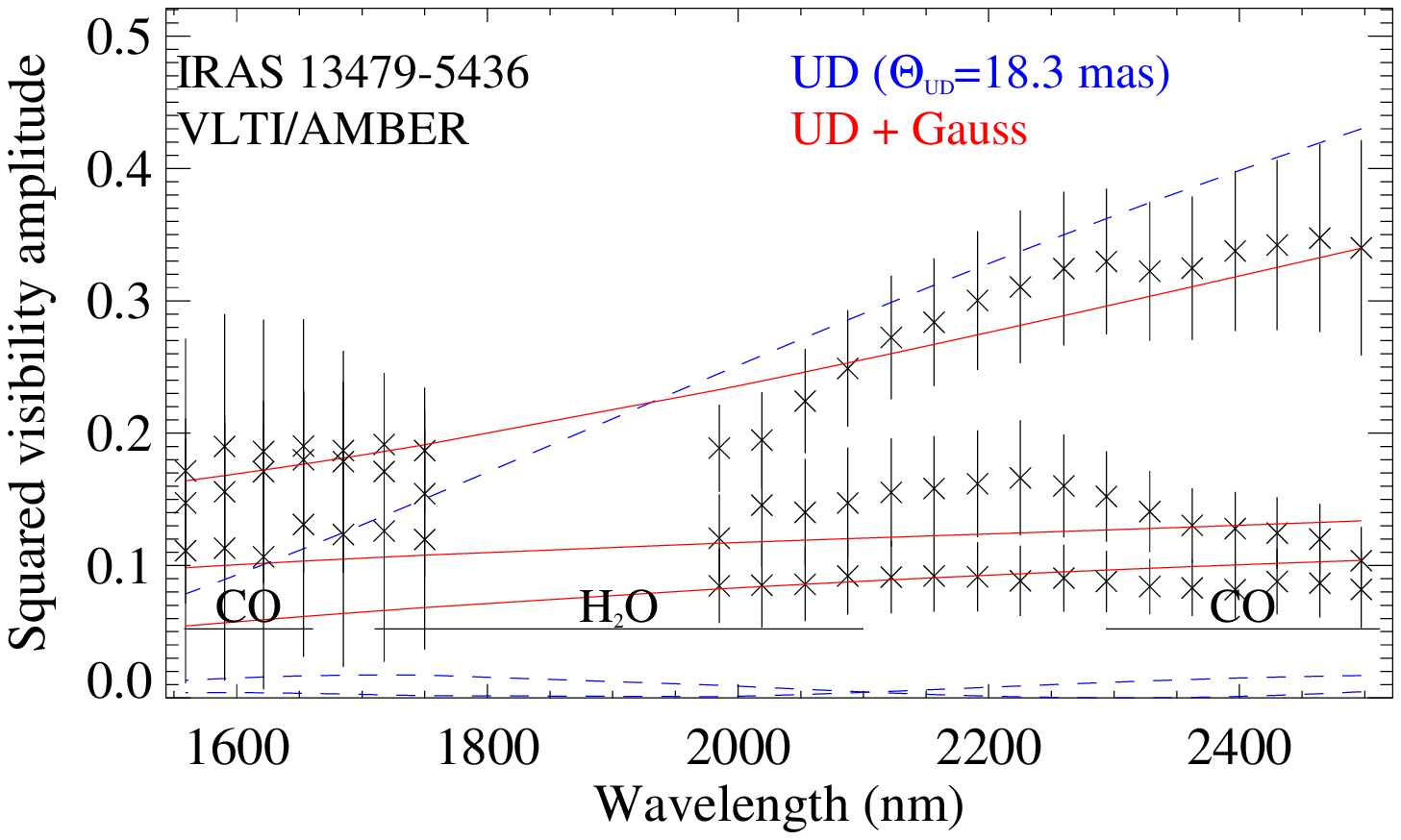}}
\resizebox{1\hsize}{!}{\includegraphics{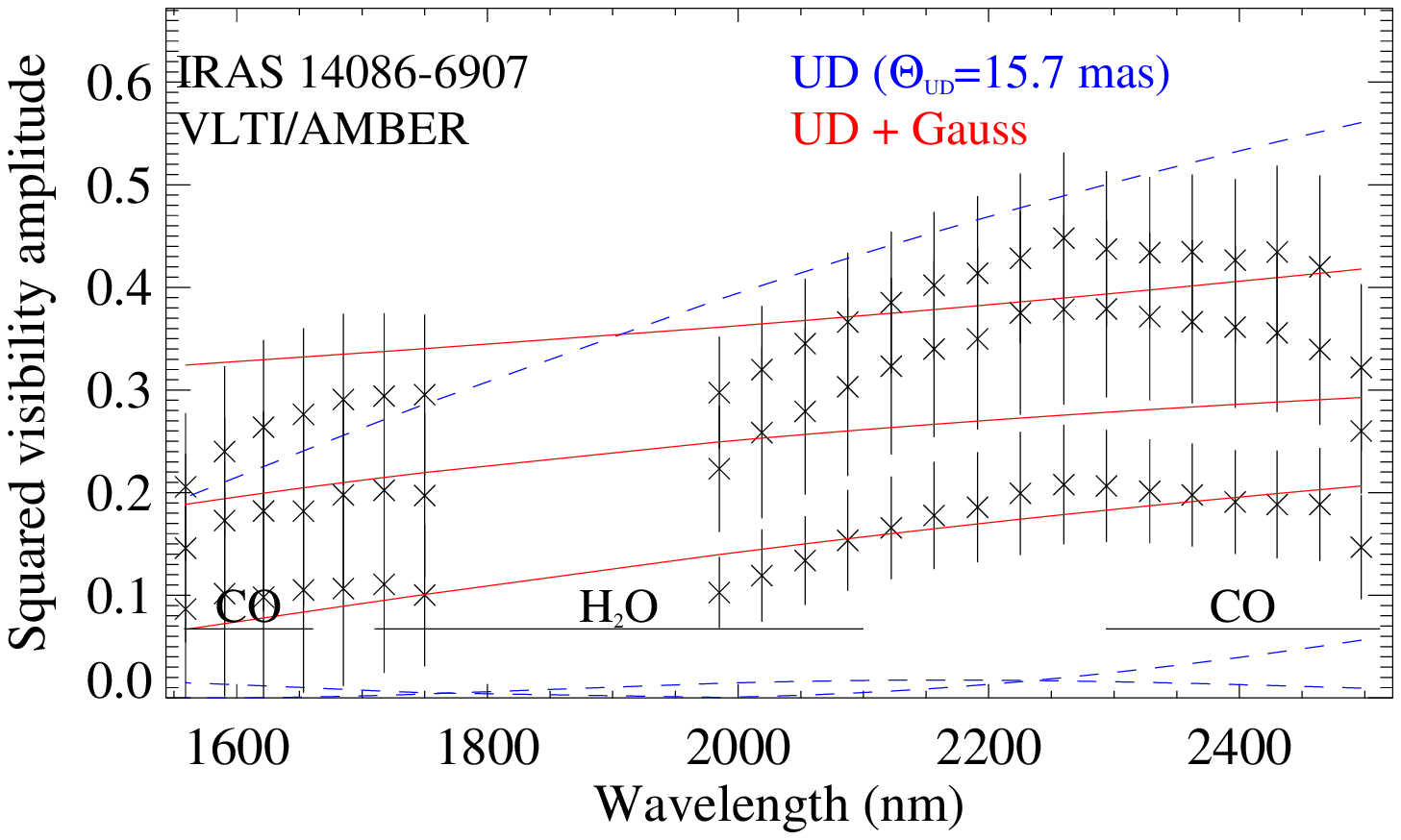}}
\resizebox{1\hsize}{!}{\includegraphics{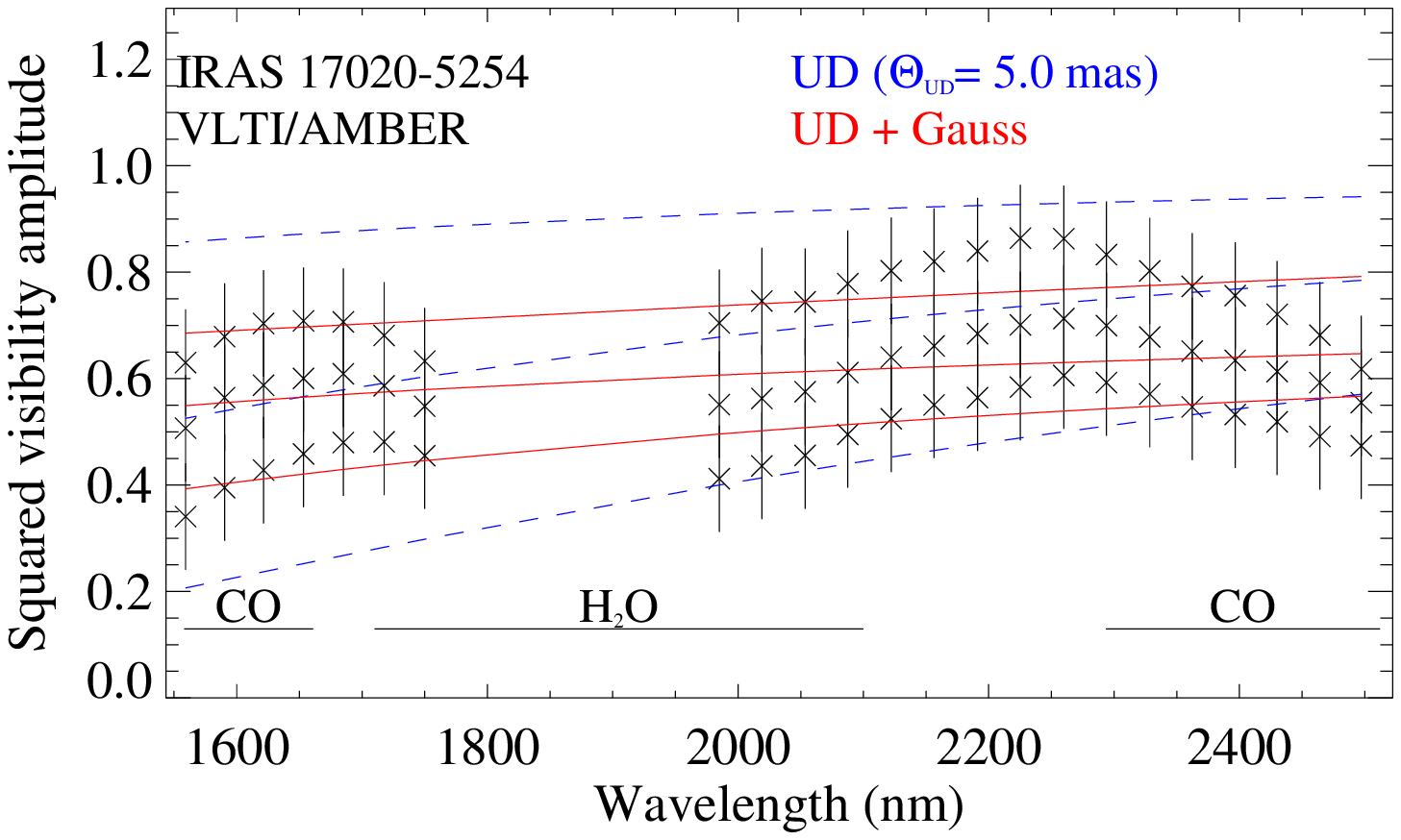}}
\caption{Squared visibility amplitudes as a function of wavelength of 
IRAS~13479-5436, IRAS~14086-6907, and IRAS~17020-5254 (from top to 
bottom). The position of H$_2$O and CO bands are indicated. 
The blue dashed lines show the best-fit of a single component uniform disk model. 
The red solid lines indicate the best-fit two-component model consisting of 
a uniform disk describing the central source and a Gaussian disk 
describing the circumstellar dust shell. The fit parameters of the 
two-component model are indicated in Tab.~\protect\ref{table:fits}.
In each panel, the three sets of lines correspond to baselines
(from top to bottom) E0-G0, G0-H0, E0-H0 with increasing projected 
baseline lengths as listed in Tab.~\protect\ref{table:log}.
}
\label{band}
\end{figure}
\begin{figure}
\centering
\resizebox{1\hsize}{!}{\includegraphics{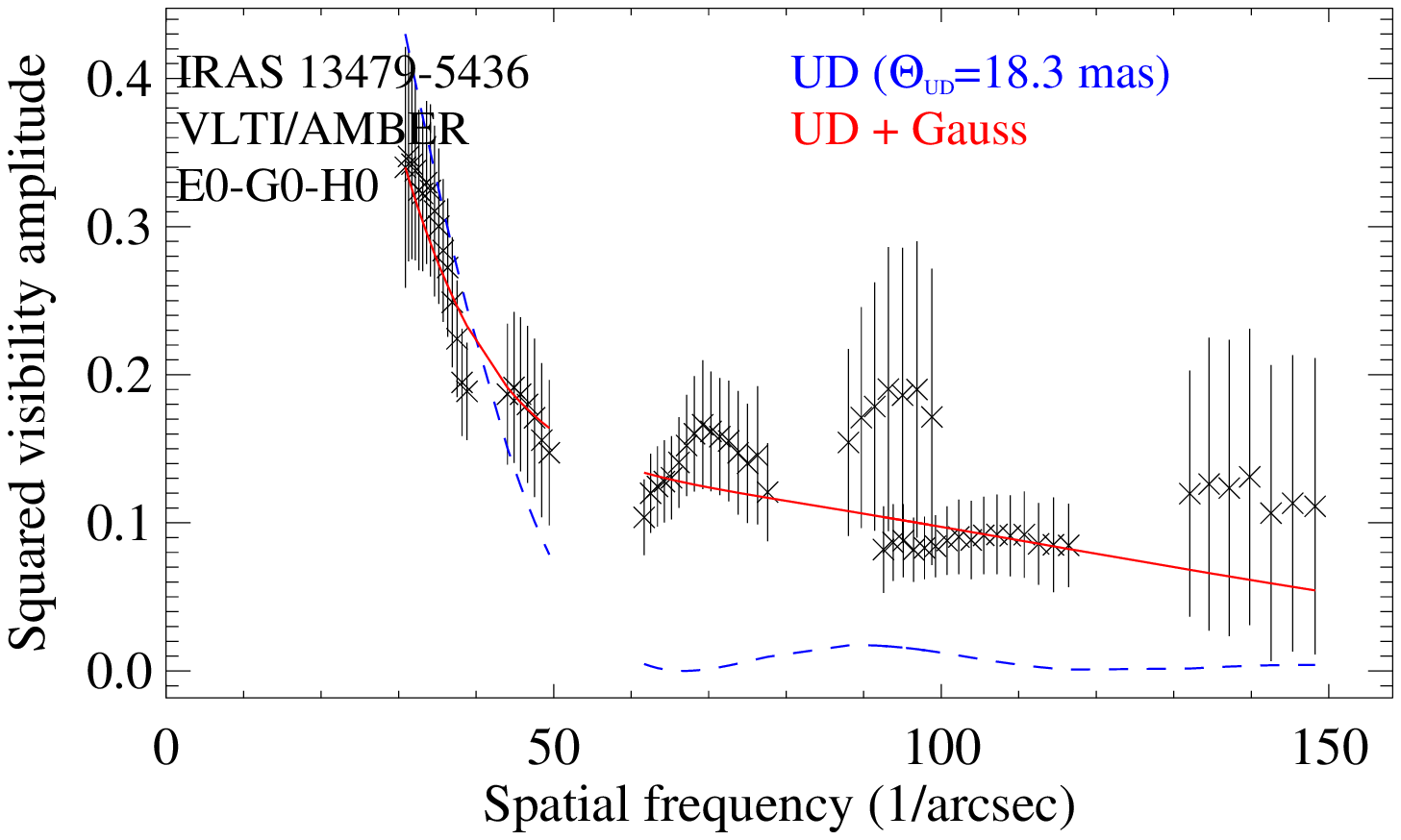}}
\resizebox{1\hsize}{!}{\includegraphics{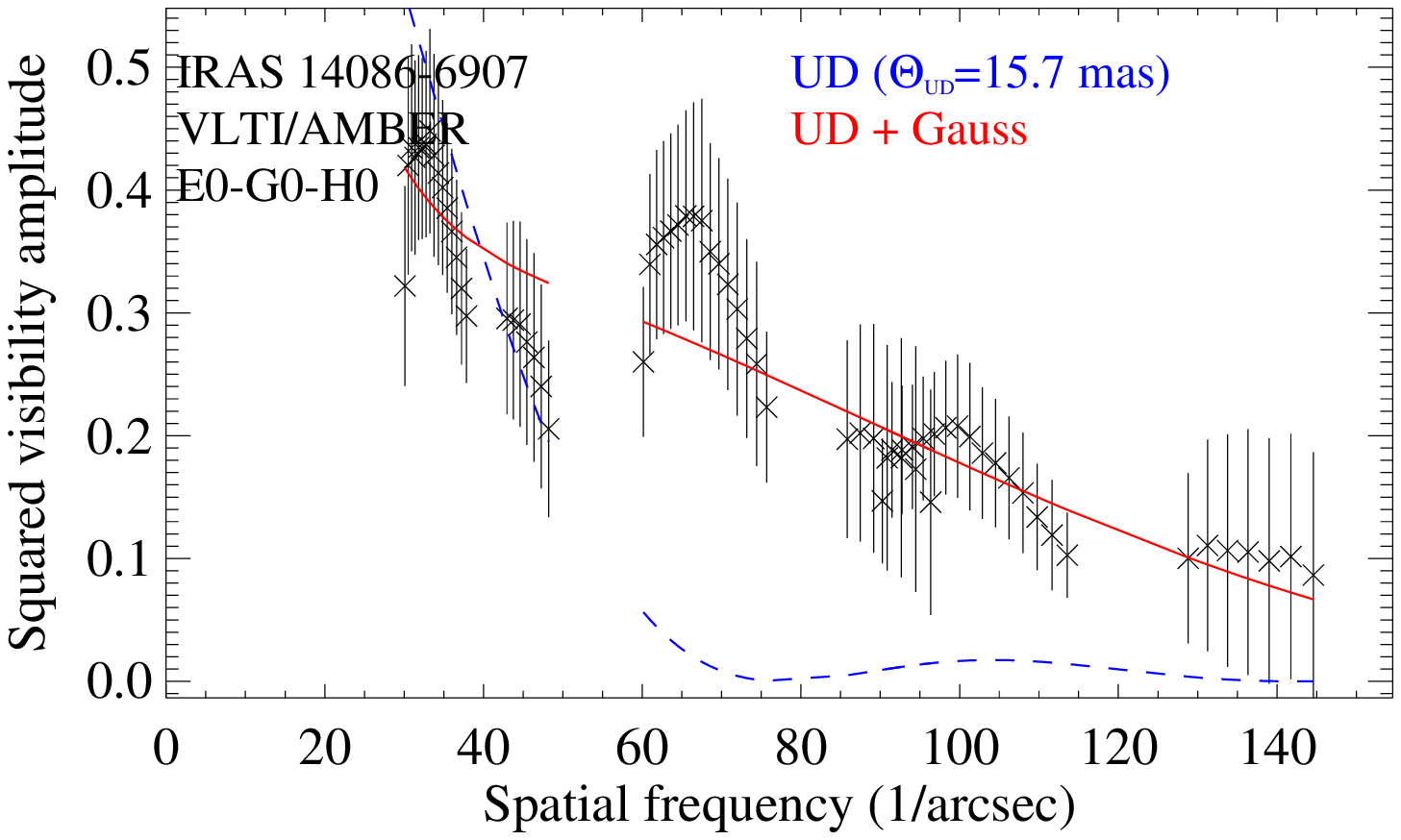}}
\resizebox{1\hsize}{!}{\includegraphics{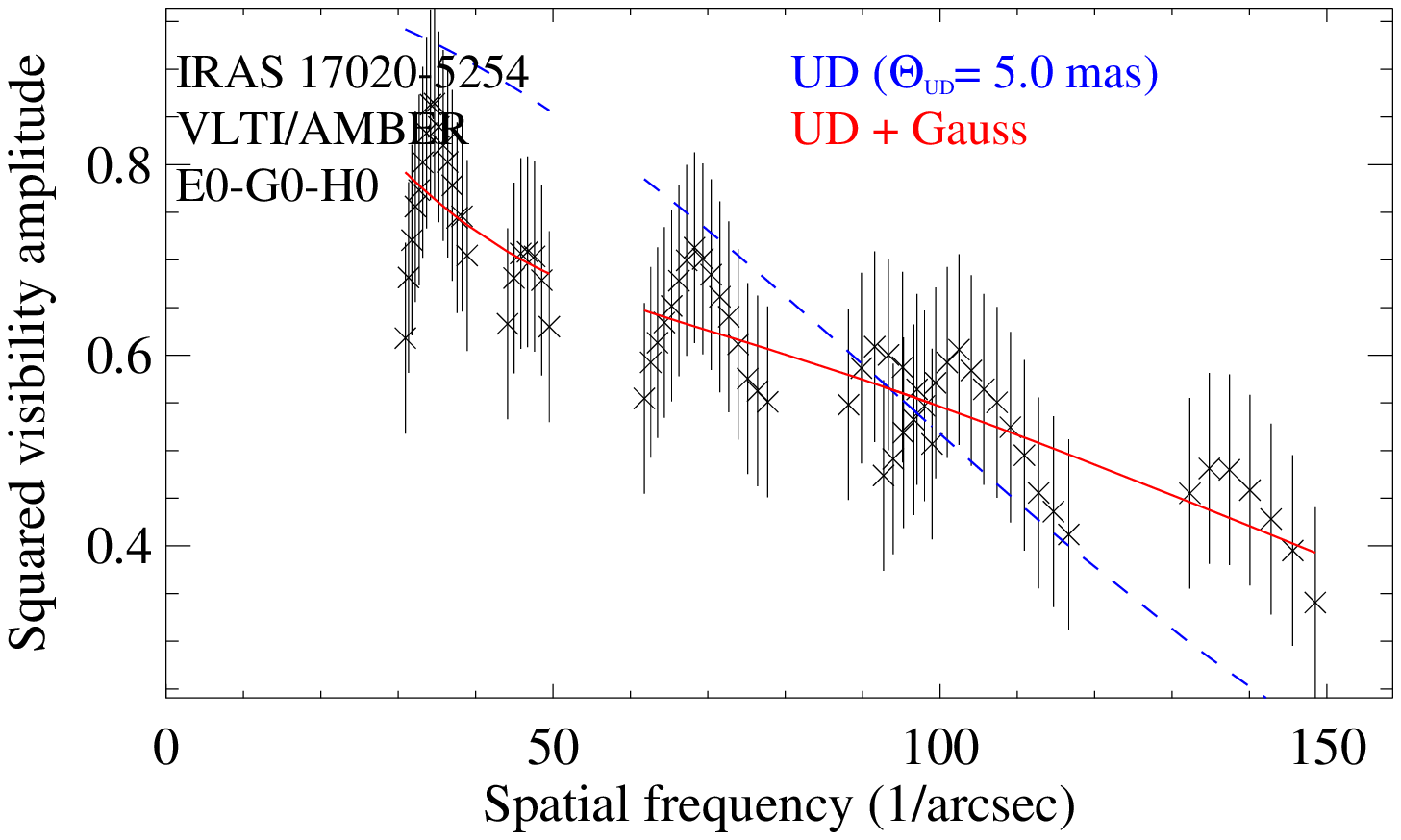}}
\caption{Squared visibility amplitudes as a function of wavelength of 
IRAS~13479-5436, IRAS~14086-6907, and IRAS~17020-5254 (from top to bottom). 
The blue dashed lines show the best-fit of a single component uniform disk model. 
The red solid lines indicate the best-fit two-component model consisting of a 
uniform disk describing the central source and a Gaussian disk describing 
the circumstellar dust shell. The fit parameters of the two-component model 
are indicated in Tab.~\protect\ref{table:fits}.}
\label{spfr}
\end{figure}

Data reduction was performed with standard procedures offered by the 
\textit{amdlib 3.0.3} package, as provided by the AMBER consortium and 
the Jean-Marie Mariotti Center \citep{tatulli07,chelli09}. 
The wavelength scale was corrected by a scaling-factor of 1.03 and 
a linear offset of -0.14\,$\mu$m. These parameters were obtained by a
comparison of the observed flux distribution of the calibrators with 
a synthetic flux curve as in \citet{wittkow11}. The synthetic flux curve
includes the telluric spectrum, the stellar spectrum, and transmission
curves of optical components of AMBER.
Visibility spectra were obtained by using an average of the two transfer 
function measurements, which were taken closest in time to the science 
target measurements. Fig.~\ref{band} shows the resulting visibility data 
as a function of wavelength and Fig.~\ref{spfr} shows the same visibility data
as a function of spatial frequency, together with visibility models
discussed below in Sect.~\ref{sec:models}.

\section{Modeling the Visibility Function}
\label{sec:models}
As a first approach to describe the visibility data, we used a 
fit of a single uniform disk (UD). A single-component model would
be expected for either a stellar source that does not show
a significant near-IR CSE, or for an optically 
thick CSE that completely obscures the central 
stellar source. The formally best-fitting curves are
indicated in Figs.~\ref{band} \& \ref{spfr} by the dashed blue
lines. Compared to the visibility data, these curves clearly
illustrate that a single-component model can not provide a 
good description of our sources. In particular, a 
single-component model can not explain the comparable small
differences of the visibility data for different baseline
lengths. 

We then used a simple geometrical two-component model of 
a UD representing the central stellar source and a Gaussian 
disk representing the dusty CSE. 
The free parameters of this two-component model are the 
angular diameter of the UD, the FWHM of the Gaussian disk,
and the flux ratio between these two components.
Following \citet{dyck2000} and \citet{berger07}, let $\theta_\mathrm{UD}$ 
be the 
angular diameter of the central UD, $\theta_\mathrm{Gauss}$ the FWHM of the
Gaussian disk, and $V_p$ the fraction of the flux contributed by the UD, i.e. 
the fraction of the attenuated stellar flux shining through the dusty envelope. 
The visibility amplitude $V(s)$ as a function of spatial frequency $s$ is then given by:
\begin{equation}
 V(s)=V_p\,\left[\frac{2\,J_1(\pi\, \theta_\mathrm{UD}\, s)}{\pi\, \theta_\mathrm{UD}\, s}\right]\, 
+\, (1-V_p)\,\left[exp\left(-\frac{(\pi\, \theta_\mathrm{Gauss}\, s)^2}{4\,ln2}\right)\right]
\end{equation}
\noindent where $J_1$ is the first-order Bessel function of the first kind.
A satisfactory fit to the visibilities was obtained using a 
Levenberg-Marquardt least-squares minimization algorithm.
The best-fit parameters of the two-component model are listed in 
Table~\ref{table:fits}. Because we use all spectral channels 
in the fit procedure, the resulting values are near-IR values
averaged over the $H-$ and $K$-bands.
\begin{table}
\centering
\caption{Fit results of the two-component model}
\label{table:fits}
\begin{tabular}{lccc}
\hline\hline
~~~~~~~IRAS  &  13479-5436      &  14086-6907     &  17020-5254\\\hline
UD diam. [mas] &  4.2 $\pm$ 0.6   &  5.4 $\pm$ 0.4  &  3.2 $\pm$ 0.4  \\
CSE, FWHM [mas]& 18.1 $\pm$ 0.9   & 25.2 $\pm$ 4.4  & 17.1 $\pm$ 3.3  \\
\% flux of UD    &  0.39 $\pm$ 0.02 & 0.62 $\pm$ 0.03 & 0.84 $\pm$ 0.02 \\
\% flux of CSE &  0.61 $\pm$ 0.02 & 0.38 $\pm$ 0.03 & 0.16 $\pm$ 0.02 \\\hline
\end{tabular}
\end{table}
The best-fit curves are indicated by the red solid lines in Figs.~\ref{band}
\& \ref{spfr}. These curves indicate that the two-component model
provides a good description of the overall shape of the visibility curves
of all three sources. 

In comparison to the two-component model, the measured visibility data
of all three sources show additional wavelength-dependent features that
are not explained by our simple UD model with wavelength-independent
diameter. 
The visibility values show
local maxima near wavelengths of 2.2\,$\mu$m in the $K$-band and near 1.7\,$\mu$m
in the $H$-band, and decreasing visibility values toward the edges of the
$H$ and $K$ bands. The decrease of the visibility values coincides with
the positions of H$_2$O and CO molecular bands, as indicated in Fig.~\ref{band}.
These visibility features resemble those that have previously been
detected in AMBER data of several Mira variable AGB stars 
\citep{wittkow08,wittkow11}, of the super-giant or super-AGB
star VX Sgr \citep{chiavassa10}, and of the semi-regular AGB star
RS Cap \citep{marti11}. These visibility features were interpreted
as being indicative of molecular layers, in the near-IR most
importantly H$_2$O and CO, lying above the continuum-forming
photosphere. In other words, this means that our simple UD 
model of the central stellar source could further be sub-divided
into a wavelength-independent photospheric continuum diameter 
and a close atmospheric molecular shell with wavelength-dependent 
molecular opacity.
The presence of the same features in the 
AMBER data of all three OH/IR stars of our sample confirms that
atmospheric molecular layers are a common phenomenon among AGB 
stars of very different luminosities and mass-loss rates, alike.
We note that the best-fit UD diameters listed in Tab.~\ref{table:fits}
are then expected to overestimate the continuum photospheric radii, 
as a result of the additional intensity contribution by the 
atmospheric molecular layers at larger radii.
For less evolved AGB stars, this overestimate may amount to up to 
about 30\%, depending on the stellar phase \citep{ireland04}.

Ideally, of course, we should match the visibility data with synthetic 
visibilities derived from physical models of OH/IR stars. However,
consistent dynamic model atmosphere and wind models are not yet available 
for oxygen-rich AGB stars, except for the attempt by \citet{jeong03}.
Also, dust-free dynamic model atmospheres do not exist for the stellar
parameters of OH/IR stars exhibiting luminosities above 10\,000\,L$_\odot$
and effective temperatures below 2000\,K \citep{lepine95}, so that
approaches of a combination of dust-free dynamic model atmospheres 
with radiative transfer models of the dust shell can not be attempted 
\citep[as performed for Mira variable AGB stars by][]{wittkow07,karovicova11}.
This means as well that our fit results of the two-component
model in Tab.~\ref{table:fits} are only indicative of the general
characteristic dimensions of the central stellar component and the 
dust shell, but do not provide accurate radius determinations of 
certain layers.

\section{Results}
\label{sec:results}
We here summarize the best-fit parameters of the above-mentioned 
two-component model.  
In order to estimate absolute sizes of the 
central stellar component, including the continuum-forming layers
and overlying molecular layers, and the outer circumstellar dust shells, 
we rely on the bolometric distances given by \citet{lepine95} 
(see Table \ref{table:colors}), albeit their uncertainties may amount 
to up to a factor of two.

\object{IRAS~13479-5436}: The best-fit visibility model consists of a
CSE with Gaussian FWHM of $18.1\pm0.9$\,mas
plus a central UD of $4.2\pm0.6$\,mas in diameter. The UD contributes 
39\% of the observed near-IR flux of this object.
The angular scales correspond to a near-IR FWHM of the CSE of 
$\sim$11300\,R$_{\odot}$ or $\sim$53\,AU and a radius of the central
stellar disk of $\sim$\,1300\,R$_{\odot}$.
For comparison, the calibration of the stellar radius as a function
of $K-L$ color by \citet{lepine95} gives a stellar radius of 
$\sim$ 1700\,R$_{\odot}$.
The rather small contribution from the stellar UD disk, as well as the 
infrared colors (largest $J-K$ color in Table~\ref{table:colors}), 
indicate a large optical depth of the CSE.

\object{IRAS~14086-6907}: The best-fit visibility model consists of a
CSE with Gaussian FWHM of $25.2\pm4.4$\,mas
plus a central UD of $5.4\pm0.4$\,mas in diameter, which contributes
62\% of the total flux. The corresponding FWHM of the CSE is
$\sim$13000\,R$_{\odot}$ or $\sim$60\,AU, and the radius of the 
central UD is $\sim$1400\,R$_{\odot}$. For comparison, the $K-L$
calibration by Lepine et al. gives a stellar radius of 
$\sim$1500\,R$_{\odot}$. 
In terms of both, stellar flux contribution and $J-K$ color, 
this object consistently takes the place between the other two.

\object{IRAS~17020-5254}: The best-fit visibility model consists of a
CSE with Gaussian FWHM of $17.1\pm3.3$\,mas
plus a central UD of $3.2\pm0.4$\,mas in diameter, contributing
84\% of the total flux. The corresponding FWHM of the CSE is
$\sim$9000\,R$_{\odot}$ or $\sim$43\,AU, and the radius of the
central UD is $\sim$860\,R$_{\odot}$ (compared 
to $\sim$912\,R$_{\odot}$ derived from the $K-L$ calibration). 
The large flux contribution from the stellar disk, as well as the small 
value of $J-K=2.4$, both indicate that we see mostly the stellar
disk including atmospheric molecular layers, with a comparably small
contribution from a diluted, fairly transparent dust-envelope. 

Please note that there is a consistent and, hence, 
convincing relation between (larger) IR-colors and (larger) flux 
contribution from the dust-rich CSE of these three objects.

\section{Discussion}
From computations of hydrodynamical wind models of long-period variables
\citep{wljhs2000} we derived CSE radii, as they would appear in the near IR, 
to compare to the observed values. Although the models available to us are 
based on a carbon-rich chemistry, we nevertheless consider this comparison to 
be of some relevance, given that both types of objects, C-rich and 
O-rich dust-enshrouded stars with large IR excess, are located at the 
tip of the AGB. In particular, we are interested in whether the 
dimensions in the O-rich case compare with those of the C-rich case, although
the details of the wind driving mechanisms are expected to be different.
We should also note that the uncertain distances 
already give this comparison a very approximate character. 

The above wind models apply 
grey radiative transfer and consequently the
resulting radii do not depend on wavelength. They vary in time, though, due 
to the stellar pulsation (prescribed by a piston approach for the inner 
boundary condition). Each wind model is characterized by a set of the 
stellar parameters mass $M$, luminosity $L$, effective temperature $T_{\rm e}$,
as well as by the carbon-to-oxygen ratio C/O of the atmospheric chemical
composition and pulsation parameters period $P$ and velocity amplitude 
$\Delta v$ of the piston. Mass-loss rates have been determined by averaging 
typically over 20 periods. 

As representative cases we selected two models \citep[w69 and w63, see 
row no. 52 and 23 of Table\,3
  in][]{wachter02} showing mass-loss rates of $1.8$ and $5.0 \times
10^{-5}\,M_\odot\,$yr$^{-1}$, respectively. The stellar parameters are 
masses of
1.0 and 0.8\,$M_\odot$, temperatures of 2800 and 2600\,K, respectively, and a
luminosity (both) of $L$=10000\,$L_\odot$. The other parameters are, in both
cases, C/O=1.3, $P$=640\,d, and $\Delta v$=5\,km\,s$^{-1}$. The observable CSE
diameter of the first model (with the lower mass-loss rate) varies between
approximately 2000 and 6000 solar radii over several pulsation cycles. The
second model (with the higher mass-loss rate) has a diameter between about 
5000\,$R_\odot$ and 9000\,$R_\odot$. With values between 
9000\,$R_\odot$ and 13000\,$R_\odot$ 
the observationally determined FWHM of the CSE of our OH/IR stars are larger 
than the CSE diameters of these models but generally of similar extension. 
The model radii based on the adopted stellar luminosity and effective temperatures
vary between 340--520\,$R_\odot$ and between 400--600\,$R_\odot$, and are 
also smaller than our best-fit UD radii and smaller than those derived
from the $K-L$ calibration by \citet{lepine95}. These differences may be 
due to different stellar parameters and/or an overestimate of the stellar
radius because of the presence of molecular layers as discussed in
Sect.~\ref{sec:models}.

>From the 1612 MHz observations \citep{lintel91} we know the expansion 
velocities of the OH maser shells. Here, the value of 
$v_{\rm exp}=17.7$\,km\,s$^{-1}$ for \object{IRAS~13479-5436} is 
larger than those of \object{IRAS~14086-6907} and 
\object{IRAS~17020-5254} (See Table~\ref{table:colors}). 
Given that the stellar luminosities are thought to increase
with increasing OH expansion velocities \citep[e.g.,][]{seven02},
\object{IRAS~13479-5436} would have a higher luminosity than
the other two sources, which is also consistent with its
largest $K-L$ color index and the calibration by \citet{lepine95}. 
Also, \citet{baudhabing} developed a relation between the 
main sequence mass and the OH outflow velocity of
$\log(M/M_{\odot})=(v_{\rm exp}-8)/16$, suggesting the same
main sequence mass of $\sim$2\,M$_\odot$ for  \object{IRAS~14086-6907} and 
\object{IRAS~17020-5254} and a larger main sequence mass
of $\sim$4\,M$\odot$ for \object{IRAS~13479-5436} together with its
higher luminosity.

Additionally, the calibrations by \citet{lepine95} as a function
of color index indicate an increasing optical depth of the CSE of OH/IR stars 
with increasing $K-L$ color index, which is supported by
our observations (see Sect.~\ref{sec:models}).
A larger optical depth of the CSE indicates a larger integrated
mass-loss history of the star. The wind models by \citet{wachter02},
which are coupled to a stellar evolutionary code, indicate that 
the maximum mass-loss rate increases with initial mass, and that
the shape becomes more spiky towards the end. Altogether we 
may speculate that the larger optical depth of the CSE of
\object{IRAS~13479-5436} compared to the other two sources
may be a result of a higher initial mass and an evolutionary phase very
close to the tip of the AGB evolution and thus the spike of the mass-loss rate. The different optical
depths of the CSEs of \object{IRAS~14086-6907} and \object{IRAS~17020-5254},
which have similar expansion velocities, may then rather be caused
by an earlier evolutionary phase of \object{IRAS~17020-5254} 
compared to \object{IRAS~14086-6907} along the same initial
mass track. Accordingly, \object{IRAS~17020-5254} could even be an 
intermediate object, lying between a Mira-variable and a tip-AGB star.

\section{Conclusions}
Using the VLTI/AMBER instrument, we determined angular diameters for 
the three OH/IR stars \object{IRAS~13479-5436}, \object{IRAS~14086-6907} and 
\object{IRAS~17020-5254} and their CSEs.
Even though our visibility fits are based on a simple two-component
geometrical model, they provide characteristic radial dimensions 
of the stars and their dust-rich shells, as well as constraints on the 
relative flux contributions from these different components. 

Our near-IR visibility data also indicate wavelength-dependent features, 
which resemble those detected in earlier AMBER observations of semi-regular 
and Mira-variable AGB stars, which are correlated with the positions
of H$_2$O and CO bands, and which are interpreted as the
result of atmospheric molecular layers lying above the continuum-forming
photospheres. This result confirms that atmospheric molecular layers
are a common phenomenon among AGB stars of very different luminosities
and mass-loss rates, alike.

We also confirm that the circumstellar dust shells contribute significantly
to the near-IR flux of our sources, unlike in the case of Mira variable
AGB stars, which can well be described by dust-free model atmospheres
at near-IR wavelengths \citep[e.g.][]{woodruff04,fedele05,wittkow08}.
This indicates a large optical depth of the CSE,
which we interpret as the result of the ``superwind'' phase, the final
10\,000 to 30\,000 years of AGB evolution, when the mass-loss rate
increases by a factor of 10-100. This result also observationally confirms 
that oxygen-rich AGB stars develop a similar ``superwind'' as carbon-rich
AGB stars, although their wind-driving mechanisms are not yet understood
to the same detail.

Even though the three stars have the same classification, interestingly
they show somewhat different characteristics. This could be attributed to 
their different initial masses and slightly different stages of evolution. 


\begin{acknowledgements}
This research has benefited from the  
\texttt{AMBER data reduction package} of the Jean-Marie Mariotti 
Center\footnote{Available at http://www.jmmc.fr/amberdrs}, and the SIMBAD 
database, operated at CDS. Strasbourg, France.
We gratefully acknowledge
support by CONACyT, Mexico and the ESO Headquarters in Garching under their
studentship programmes (ARV), and by PROMEP, Mexico, project 103.5/10/4684 (AW).
\end{acknowledgements}

\end{document}